\documentclass[aps,prd,twocolumn,showpacs,superscriptaddress]{revtex4-1}
\usepackage{graphics}
\usepackage{dcolumn}
\usepackage{bm}
\usepackage{mathrsfs}
\usepackage{pstricks}
\usepackage{color}
\usepackage{slashed}
\usepackage{amsmath}
\usepackage{epsfig}
\usepackage{amsfonts}
\usepackage{amssymb}
\def\beq{\begin{equation}}
\def\eeq{\end{equation}}
\def\bea{\begin{eqnarray}}
\def\eea{\end{eqnarray}}
\def\nn{\nonumber}

\def\x{\mathbf{x}}

\def\p{\mathbf{p}}

\begin{document}

\title{Relativistic Bose-Einstein condensation with disorder}
\author{E. Arias}
\email{earias@if.uff.br}
\affiliation{Instituto de F\'{\i}sica, Universidade Federal Fluminense\\
Av. Gal. Milton Tavares de Souza -- Campus da Praia Vermelha, 24210-346  Niter\'oi, RJ, Brazil}
\author{G. Krein}
\email{gkrein@ift.unesp.br}
\affiliation{Instituto de F\'\i sica Te\'orica, Universidade Estadual Paulista\\
Rua Doutor Bento Teobaldo Ferraz 271 - Bloco II, 01140-070  S\~ao Paulo, SP, Brazil}
\author{G. Menezes}
\email{gabrielmenezes@ufrrj.br}
\affiliation{Departamento de F\'isica, Universidade Federal Rural do Rio de Janeiro\\
BR 465--07, 23890--971 Serop\'edica, RJ, Brazil}
\author{N. F. Svaiter}
\email{nfuxsvai@cbpf.br}
\affiliation{Centro Brasileiro de Pesquisas F\'{\i}sicas, Rua Dr. Xavier
Sigaud 150, 22290-180  Rio de Janeiro, RJ, Brazil}
%

\begin{abstract}
We investigate the thermodynamics of a self-interacting relativistic charged scalar
field in the presence of weak disorder. We consider quenched disorder which couples linearly to 
the mass of the scalar field. After performing noise averages over the free energy of the system, 
we find that disorder increases the mean-field critical temperature for Bose-Einstein condensation 
at finite density. The effect of disorder on the temperature dependence of the chemical 
potential for a fixed charge density is investigated. Significant differences from the mean-field
temperature dependence of the chemical potential are observed as the strength of the 
noise intensity increases. Finally, the temperature dependence of the chemical potential with fixed total
charge and entropy is investigated. It is found that there is no Bose-Einstein condensation for
a fixed charge to entropy ratio in the presence of weak disorder. The possible relevance of 
the findings in the present paper in different areas is discussed. 
\end{abstract}

\pacs{11.10.-z, 03.70.+k, 05.30.Jp, 05.70.Fh,05.40.-a,42.25.Dd}

\maketitle

\section{Introduction and motivation}
\label{intro}

Disorder plays an important role in the critical behavior of second order
phase transitions~\cite{dot}. The relevance of disorder in the criticality
can be assessed qualitatively using the critical exponent  $\alpha$ of the
specific heat for the disorder-free system~\cite{harris,harris2}; namely, when
$\alpha > 0$ (the specific heat diverges at the critical point), the
critical behavior of the disordered system is changed, when
$\alpha < 0$ (the specific heat is finite), disorder has no effect on
the critical behavior. On the other hand, at low temperatures quantum
fluctuations may compete with the random fluctuations; an example is the
destruction of the ordered ground state of a spin-glass -- a disorder
strongly correlated system -- by quantum fluctuations~\cite{sachdev}.
Conversely, quantum fluctuations can stabilize a glass phase in a disordered
environment; Carleo~{\it et~al.}~\cite{carleo} demonstrated that repulsively
interacting bosons can feature a novel quantum phase displaying both
Bose-Einstein condensation and spin-glass behavior due to frustration.
In the present paper we investigate the interplay between quantum and
random fluctuations in a self-interacting relativistic charged scalar field
theory with a finite chemical potential. Disorder in relativistic Bose-Einstein
condensation has not been considered in the literature, contrary to the case of
non-relativistic Bose-Einstein condensation, where it has been under
intensive study since early seminal works~\cite{{hertz},{huang},{fisher}}.

Disorder has a decisive influence on the zero-temperature phase diagram of
non-relativistic Bose systems. As emphasized by the literature, there is a
quantum phase transition for such systems from a Mott insulating phase to a
conducting phase. Since no pure Bose system can be a normal conducting fluid at zero
temperature, the conducting-insulator transition must correspond to the
onset of superfluidity. As shown in Ref.~\cite{fisher}, this scenario is changed
dramatically in the presence of a random potential. For the case of a Gaussian
colored noise, a Bose glass phase also arises and the transition to superfluidity
only occurs from this third phase, never directly from the Mott insulator. The
introduction of a random potential in such systems may also imply the destruction
of the superfluidity phase, as discussed in Refs.~\cite{{huang},{dis2a,dis2b,dis2c,dis2d}}. In particular,
a recent study by Lopatin and Vinokur~\cite{lopatin} employing the replica method
found a negative shift in the condensation temperature of a dilute Bose gas due to
disorder -- see also Refs.~\cite{kobayashi,falco}.

There is an extensive literature on relativistic Bose-Einstein condensation (RBEC)
following the pioneering works of Refs.~\cite{{karsh},{aragao},{kapustax},{haber,haber2},
{frota},{bernstein,bernstein2}}, which discussed RBEC in flat space-times, and Refs.~\cite{{singh},
{parker},{shiraishi},{toms}}, which discussed RBEC in curved space-times. While relativistic
Bose-Einstein condensates are not yet realizable in controllable experiments like their
non-relativistic counterparts, they do relate to observable and experimentally accessible
phenomena. One example, of immense current interest, concerns the condensation dynamics
in relativistic quantum field theories where creation and annihilation of particles play
crucial role, like in far-from-equilibrium stages of the early Universe and in experiments
with relativistic heavy-ion collisions~\cite{berges}. There is also the possibility of
Bose-Einstein condensation of pions and kaons~\cite{{bed-schaf},{bedaque}, {kap-red},{andersen}}
in neutron stars. The condensation of these mesons will affect the equation of state of matter
in the interior of the star, which has direct consequences on the observable mass-radius
relation of the star, and will also impact the early evolution of the neutron star.
In turn, in dark-matter models where scalar particles constitute a natural ingredient,
relativistic Bose-Einstein condensates assume an important place in the study of the
effects of scalar dark-matter background on the equilibrium of degenerate stars~\cite{grifols}.
In this case there is particular interest in the charge density and the associated
chemical potential.

In real physical situations, the presence of some sort of disorder in the system is unavoidable. The disorder
can be due to uncontrollable disturbances external to the system; for instance in a cosmological context such
perturbations can originate from standard inflationary fluctuations, required to generate large-scale structures.
On the other hand, random fluctuations can also be the result of an incomplete treatment of degrees of freedom
associated with fields that couple to the field of interest. As with nonrelativistic Bose-Einstein condensates
of condensed matter physics, one expects that disorder will impact the critical behavior of relativistic
Bose-Einstein condensation. The present study is a first step toward a systematic study of disorder in
relativistic quantum field theory models, in that we focus on a weakly interacting charged scalar field at
finite temperature in the presence of nonstatic randomness (the precise meaning for {\em nonstatic noise} will
be defined shortly). Our model is a kind of generalization of the scalar Landau-Ginzburg theory, where the
quenched disorder is described by random fluctuations of the effective transition temperature~\cite{dot}.

The organization of this paper is as follows. In Sec.~II we present our model. The disorder field couples to
the charged scalar field via the mass term of the scalar field, just as in the random-temperature Landau-Ginzburg model.
We consider weak disorder and implement a perturbative expansion for the free energy as power series expansion in the
strength of the disorder field. In Sec.~III we study the thermodynamics properties of the self-interacting relativistic
Bose gas at finite density with randomness. The self-interactions of the scalar field are treated in a mean-field
approximation. We calculate the noise average of the free energy. In Sec.~IV we obtain the critical temperature in
the presence of random fluctuations. In Sec.~V we discuss the net total charge associated with the condensate and also
the modifications in temperature evolution of the chemical potential due to disorder. Conclusions and Perspectives are
presented in Sec.~VI. The paper includes Appendices containing details of lengthy derivations. Throughout the paper
we employ units with $\hbar=c=k_{B}=1$.

\section{Scalar field thermodynamics and disorder}
\label{model}

We are interested in studying the effects of randomness on a charged scalar
field $\varphi$ of mass~$m$ in equilibrium with a thermal reservoir
at temperature~$T$. We employ the imaginary time formalism of Matsubara~\cite{mats}
to write the partition function of the model in the grand canonical ensemble
as~\cite{kapustax,bernard}
\begin{equation}
Z=\left[N(\beta)\right]^2\,\int [D\varphi][D\varphi^{*}]\,\,e^{S[\varphi,\varphi^{*}]},
\label{50}
\end{equation}
where the action $S[\varphi,\varphi^{*}]$ reads
\bea
S[\varphi,\varphi^{*}] &=&  \int_0^\beta d\tau\,\int_{V} d\x\,\bigg[(\partial_{t}+i\mu)\,
\varphi^*\,(\partial_{t} - i\mu)\,\varphi
\nn\\
&&
\,- \nabla\varphi^{*}\nabla\varphi - m^2\varphi^*\varphi - \lambda(\varphi^*\varphi)^2\bigg],
\label{ss0}
\eea
where $V$ is the volume of the system, $\beta = 1/T$, $\mu$ the chemical potential
associated with the conserved charge, and $\partial_t = i\,\partial_{\tau}$.
The field $\varphi$ satisfies the Kubo-Martin-Schwinger~\cite{mks,mks2} boundary condition
$\varphi(\tau,\x) = \varphi(\tau+\beta,\x)$. $N(\beta)$ is a $\beta$-dependent but
$\mu$-independent constant that comes from the integration over the canonical momentum
conjugated to the field $\varphi$~\cite{bernard}.

Next we consider the coupling of a random noise source to the quantum matter field
in a similar fashion to the random-temperature Landau-Ginzburg model, but generalized
to a $\tau$-dependent noise. That is, we perform the replacement $m^2\rightarrow m^2(1+\nu)$,
where $\nu =\nu(\tau,\x)$ is dimensionless. The partition function given in
Eq.~(\ref{50}) becomes replaced by
\begin{equation}
Z[\nu] = \left[N(\beta)\right]^2 \, \int [D\varphi][D\varphi^{*}] \;
e^{S_{T}[\nu,\varphi,\varphi^{*}]},
\label{z-random}
\end{equation}
where
\begin{equation}
S_ {T}[\nu,\varphi,\varphi^{*}] = S[\varphi,\varphi^{*}] + S_I[\nu,\varphi,\varphi^{*}],
\label{total-action}
\end{equation}
with $S[\varphi,\varphi^{*}]$ given by Eq.~(\ref{ss0}) and $S_I[\nu,\varphi,\varphi^{*}]$
contains the coupling of the scalar field with the noise field:
\begin{equation}
S_I[\nu,\varphi,\varphi^*] = - m^2\int_0^\beta d\tau\int_{V} d\x \,
\nu(\tau, \x)\varphi^*(\tau, \x) \varphi(\tau, \x).
\label{s-random}
\end{equation}
The physical picture is that the random fluctuations describe average effects of external
disturbances on the system or of degrees of freedom of unobserved fields. Although similar
to a real-time dependence, the $\tau$ dependence in $\nu(\tau,\x)$ should be understood
as being of similar nature of the one that arises naturally in a self-energy for the
field $\varphi$ when integrating out fields in favor of effective interactions of $\varphi$.
It is important to note that in general, when integrating over unobserved degrees of
freedom one obtains also effective vertices, in addition to self-energies. Thereof we
stress that there is no implicit assumption here that Eq.~(\ref{s-random}) is an exact
replacement for all effects of integrating out unobserved fields, but solely that the
dependence on $\tau$ of the noise field is very natural for non-isolated systems. Hereafter we mean by {\em static noise} the noise fields that are $\tau$ independent and {\em nonstatic noise} those fields that depend
upon $\tau$. Reference~\cite{time} presents another situation in which the noise is nonstatic.

Here we consider the random function $\nu(\tau, \x)$ as a Gaussian distribution given by
\begin{equation}
P[\nu]=p_{0}\,e^{-1/2\sigma^2 \int\,d^{d}x \, \left[\nu(x)\right]^2 }
\label{dis2}
\end{equation}
where $x = (\tau, \x)$ and $p_{0}$ is the normalization constant of the distribution.
The quantity $\sigma^2$ is a parameter associated with the intensity of the disorder.
We will denote the mean value over the random variable as $\overline{(\cdots)}$,
defined by
\begin{equation}
\overline{ A[\nu] } = \int[D\nu]\,P[\nu]\,A[\nu],
\label{aveA}
\end{equation}
with $A[\nu]$ being any functional of $\nu$. From Eq.~(\ref{dis2}), we have a white noise with
two-point correlation function given by
\begin{equation}
\overline{\nu(\tau, \x)\nu(\tau', \x')} = \sigma^2\,\delta(\tau-\tau')\delta^{3}(\x-\x').
\label{corr}
\end{equation}
As well known, it follows from the Gaussian distribution that
\begin{equation}
\overline{ \nu(x_1)\cdots\nu(x_{2n+1}) } =0,
\label{med}
\end{equation}
\begin{eqnarray}
\overline{ \nu(x_1)\cdots\nu(x_{2 n}) }\, &&=\sum_{pair\,comb.}
\prod_{pairs}\overline{\nu(x_j)\nu(x_k)},
\end{eqnarray}
where $n$ is an integer.

The standard procedure to study Bose-Einstein condensation is to separate from $\varphi$
the constant zero mode $\langle \varphi \rangle \equiv \xi$:
\begin{equation}
\varphi=\xi+\chi,
\label{mu}
\end{equation}
where $\chi$ is a complex field with no zero mode. The $\chi$ field is written in terms of
real and imaginary parts as
\begin{equation}
\chi=\frac{1}{\sqrt{2}}(\chi_1+i\chi_2) ,
\label{sep}
\end{equation}
so that the action in Eq.~(\ref{total-action}) can be written as
\bea
S_T[\nu, \chi_1,\chi_2,\xi] &=& -\beta V\,U(\xi) + S_0[\xi,\chi_1,\chi_2]
\nn\\
&+&S_{int}[\xi,\chi_1,\chi_2] + S_I[\nu, \chi_1,\chi_2,\xi],
\eea
with the following potential
\begin{equation}
U(\xi)=(m^2-\mu^2)\xi^2 + \lambda\xi^4,
\label{class}
\end{equation}
the quadratic part
\begin{eqnarray}
S_0[\xi,\chi_1,\chi_2] &=& -\frac{1}{2}\int_0^\beta d\tau\int_{V} d\x\,
\bigg[\partial_\tau\chi_1\partial_\tau\chi_1 +\nabla\chi_1\nabla\chi_1
\nn\\
&+& (6\lambda\xi^2 + m^2-\mu^2)\chi_1^2 + \partial_\tau\chi_2\partial_\tau\chi_2
\nn\\
&+&  \nabla\chi_2\nabla\chi_2
+ (2\lambda\xi^2 + m^2-\mu^2)\chi_2^2
\nn\\
&-&2i\mu(\chi_2\partial_\tau\chi_1 - \chi_1\partial_\tau\chi_2)
\bigg],
\label{s0}
\end{eqnarray}
and the self-interacting part
\bea
S_{int}[\xi,\chi_1,\chi_2] &=& -\int_0^\beta d\tau\int_{V} d\x\,\bigg[2^{1/2}\lambda\xi\chi_1\left(\chi_1^2 + \chi_2^2\right)
\nn\\
&+& \frac{\lambda}{4}\left(\chi_1^2 + \chi_2^2\right)^2\bigg].
\label{s-self}
\eea
In Eqs.~(\ref{s0}) and~(\ref{s-self}) we neglected the linear terms in the field $\chi$, because their contributions
will be proportional to terms like $\chi(\p = 0)=0$. In turn, the random contribution is given by
\bea
S_I[\nu, \chi_1,\chi_2,\xi] &=& -m^2\int_{0}^{\beta}d\tau\int_{V}d\x\,\biggl\{\xi^2\,\nu(\tau,\x)
\nn\\
&-& \frac{1}{2}\,\nu(\tau,\x)\Bigl[\chi^2_1(\tau,\x) + \chi^2_2(\tau,\x)\Bigr]
\nonumber\\
&-& \sqrt{2}\,\xi\,\nu(\tau,\x)\chi_1(\tau,\x)\biggr\}.
\label{si-random}
\eea
We are interested in studying the thermodynamics of the above system in the presence of disorder. We follow closely
the path used for the noiseless case~\cite{kapustax,haber,haber2,kapusta}, in that the transition temperature is determined
by analyzing the minimum of the free energy as a function of the variational parameter~$\xi$. Noise average is taken
into account using Eq.~(\ref{aveA}), with $A$ being the Helmholtz free energy $\Omega(\beta, V, \mu, \xi)$. Specifically,
for a uniform infinite volume system we have the relation $\beta \Omega(\beta, V, \mu, \xi) = -\ln Z_R$, where $\ln Z_R$
is the renormalized logarithm of the partition function, and thence:
\bea
\overline{\Omega}(\beta, V, \mu, \xi) &=& -\frac{1}{\beta}\int[D\nu]\,P[\nu]\; \ln Z_R[\nu]
\nn\\
&=& -\frac{1}{\beta}\,\overline{\ln Z_R[\nu]}.
\label{17}
\eea
Note that we are considering a situation where one has to deal with two kinds of averages,
namely thermal averages and noise averages, which are not treated on the same footing.
This can be justified when the characteristic time scale of the change in disorder is much
larger then the time of observation of phenomena of interest. This means that in order to
calculate random averages of thermodynamic observables, one performs such averages
{\em over the logarithm of the partition function} and not over the partition function
itself. The noise average over the partition function is trivial, as one can integrate
very easily over $\nu(x)$ using the probability distribution of Eq.~(\ref{dis2}). In other words,
one calculates the free energy for a given configuration of the noise $\nu(x)$ and then
carry out the random average.

Eq.~(\ref{17}) requires a method to evaluate the average over noise realizations of the
free energy. For static noise and arbitrary noise intensities the replica-trick is widely
used~\cite{dot}. Here we consider the weak-noise limit and use a perturbative
approach~\cite{krein1,new}, in that one expands the partition function in a power series
in the noise~$\nu$. This will be discussed in the next Section.

\section{Noise average of the free energy}
\label{logZ}

It is known that random mass models generate effective interactions that mimic a negative coupling
constant. Because of this, we will consider the mean field approximation for the disorder-free part of the
partition function; i.e. one calculates the noiseless free energy neglecting $S_{int}[\xi,\chi_1,\chi_2]$.
As discussed in Ref.~~\cite{kapusta}, one might expect this to be a good approximation if both $
\lambda$ and $\lambda\xi$ are small. We note that there is no assumption here that a mean field approximation
captures the full richness of the critical behaviour of the relativistic interacting Bose gas; the approximation
is used because it provides the system with a ground state and a starting point for assessing the role played by
disorder in the relativistic model. Therefore, the model only makes sense when the noise-induced interactions are
weaker than the self-interactions $\lambda\,(\varphi^*\varphi)^2$ in Eq.~(\ref{ss0}). In our treatment we
ensure this by treating the noise as a weak interaction on the top of the mean-field generated by the
self-interactions $\lambda\,(\varphi^*\varphi)^2$. In other words, the noise is weakly coupled to the scalar
field in such a way that the random fluctuations do not destabilize the mean field solution and still allows
for the existence of a ground state.

In the weak-disorder limit, the partition function in Eq.~(\ref{z-random}) can be
expanded in a power series in $S_I$:
\begin{equation}
Z[\nu]=(N(\beta))^2\int[d\chi_1]\,[d\chi_2]e^{-\beta V\,U(\xi) + S_0} \,
\sum_{n=0}^{\infty}\,\frac{S_I^n}{n!},
\end{equation}
where $S_I=S_I[\nu,\chi_1,\chi_2,\xi]$ and $S_0 = S_0[\xi,\chi_1,\chi_2]$. Taking the
logarithm of both sides and then taking the random average leads us to
\begin{equation}
\overline{\ln Z[\nu]} = \ln Z_{MF} + \overline{\ln Z_I[\nu]},
\label{helm}
\end{equation}
where the mean-field contribution $\ln Z_{MF}$  is given by $\ln Z_{MF} = - \beta V\,U(\xi)
+ \ln Z_0$, with
\begin{equation}
\ln Z_0 = \ln\biggl[(N(\beta))^2\,\int [D\chi_1][D\chi_2]\,e^{S_{0}}\biggr].
\label{z0}
\end{equation}
The quantity $\ln Z_0$ is calculated explicitly in Appendix~\ref{app1} and the result is
\bea
\ln Z_0 &=&  - \sum_{\p}\bigg[\frac{\beta\Omega_{+}}{2}+\ln\left(1-e^{-\beta\Omega_{+}}\right)
\nn\\
&+&\frac{\beta\Omega_{-}}{2} +\ln\left(1-e^{-\beta\Omega_{-}}\right)\bigg],
\label{a10}
\eea
where the quantities $\Omega_{\pm}$ are properly defined in the Appendix~\ref{app1}. Therefore, one gets the following
mean-field partition function:
\bea
\ln Z_{MF} &=& - \beta V\,U(\xi) - \sum_{\p}\bigg[\frac{\beta\Omega_{+}}{2}+\ln\left(1-e^{-\beta\Omega_{+}}\right)
\nn\\
&+&\frac{\beta\Omega_{-}}{2} +\ln\left(1-e^{-\beta\Omega_{-}}\right)\bigg].
\label{mf}
\eea
Now let us focus on the corrections to the mean-field solution due to disorder which are given by
\begin{equation}
\overline{\ln Z_I[\nu]} = \overline{\ln\biggl(1 + \sum_{n=1}^{\infty}\frac{\langle S_{I}^n
\rangle}{n!}\biggr)}.
\label{zi}
\end{equation}
Here the averages $\langle(\cdots)\rangle$ are defined using the mean-field ensemble represented by the action $S_0$:
\begin{equation}
\langle(\cdots)\rangle = \frac{\int[D\chi_1][D\chi_2](\cdots)\,e^{S_0}}
{\int[D\chi_1][D\chi_2]\,e^{S_0}}.
\label{51}
\end{equation}
Expanding Eq.~(\ref{zi}) up to second order in the noise field, one obtains
\begin{equation}
\overline{\ln Z_I[\nu]} = \overline{\langle S_{I}\rangle}+\frac{1}{2}
\bigg(\overline{\langle S_{I}^2\rangle} - \overline{\langle S_{I}\rangle^2}\bigg),
\end{equation}
where $S_I$ is given by Eq.~(\ref{si-random}). From Eq.~(\ref{med}), we have that $\overline{\langle S_I\rangle}=0$.
The other terms are obtained using Eqs.~(\ref{si-random}) and~(\ref{corr}):
\bea
\overline{\ln Z_I[\nu]} &=&  m^4\sigma^2 \int_0^\beta d\tau\int_{V} d{\x}
\Biggr[\,\xi^2\,\bigl\langle \chi_1^2 \bigr\rangle + \frac{1}{8}\,
\biggl(\bigl\langle \chi_1^4 \bigr\rangle
\nn\\
&+&\bigl\langle \chi_2^4 \bigr\rangle
- \bigl\langle \chi_1^2\bigr\rangle^2 - \bigl\langle \chi_2^2 \bigr\rangle^2
\nn\\
&+& 2\,\bigl\langle \chi_1^2 \,\chi_2^2 \bigr\rangle - 2\,\bigl\langle
\chi_1^2 \bigr\rangle\bigl\langle \chi_2^2 \bigr\rangle\biggr)\Biggr].
\label{zi-ave}
\eea
The derivation of the ensemble averages in Eq.~(\ref{zi-ave}) can be performed in the usual way
(see for instance Ref.~\cite{kapusta}). The result is
\bea
\overline{\ln Z_I[\nu]} &=& m^4\sigma^2\Biggl[\xi^2
\sum_{n,\p}\,{\cal D}^0_{11}(\omega_n,\p)
\nn\\
&+& \frac{1}{4\,\beta V}\,\left(\sum_{n,\p}\,
{\cal D}^0_{11}(\omega_n,\p)\right)^2\,
\nn\\
&+& \frac{1}{4\,\beta V}\,\left(\sum_{n,\p}\,
{\cal D}^0_{22}(\omega_n,\p)\right)^2\,\Biggr],
\label{zi-ave1}
\eea
where ${\cal D}^0_{ij}(\omega_n,\p)$, $i,j=1,2$ are the zero-order propagators
of the fields $\chi_j$. Since the propagators have divergent vacuum contributions,
Eq.~(\ref{zi-ave1}) must be carefully regularized. The renormalization of the
propagators is discussed in Appendix~\ref{app3}. After carrying out such a procedure
we get
\bea
\overline{\ln Z_I[\nu]} = (\beta V)\frac{m^4\sigma^2}{2}\Bigl[\Pi_m(2\xi^2 + \Pi_m) - \Pi_v^2\Bigr],
\label{ln-ren}
\eea
where the quantities $\Pi_v$ and $\Pi_m(\beta, \xi)$ are obtained in Appendix~\ref{app3};
they are given by
\beq
\Pi_v = \frac{1}{4}\int\frac{d\bf{p}}{(2\pi)^3}\frac{1}{W_{+}(\p,\xi)}
+ \frac{1}{4}\int\frac{d\bf{p}}{(2\pi)^3}\frac{1}{W_{-}(\p,\xi)},
\eeq
and
\bea
\Pi_m &=& \Pi_m(\beta, \mu, \xi) = \frac{1}{2}\int\frac{d\bf{p}}{(2\pi)^3}\frac{1}{W_{+}(\p,\xi)}
\frac{1}{e^{\beta\Omega_{+}}-1}
\nn\\
&+&\frac{1}{2}\int\frac{d\bf{p}}{(2\pi)^3}\frac{1}{W_{-}(\p,\xi)}\frac{1}{e^{\beta\Omega_{-}}-1}.
\label{pim-30}
\eea
The quantities $W_{\pm}(\p,\xi)$ are properly defined in Appendix~\ref{app3}. Note
that in the above equations we have considered the large-volume limit. Finally,
inserting Eqs.~(\ref{class}),~(\ref{mf}) and~(\ref{ln-ren}) in equation~(\ref{helm}) and
neglecting for the moment the divergent vacuum contributions, one gets the following
expression for the renormalized $\ln Z$ up to second order in noise intensity:
\bea
\frac{1}{\beta V}\,\overline{\ln Z_R[\nu]} &=&  \,\left[\mu^2 - m^2 - \lambda\xi^2
+ m^4\sigma^2\,\Pi_m\right]\xi^2
\nn\\ 
&-& \frac{1}{\beta}\int\frac{d{\bf p}}{(2\pi)^3}\biggr[\ln\left(1-e^{-\beta\Omega_{+}}\right)
\nn\\
&+&\ln\left(1-e^{-\beta\Omega_{-}}\right)\biggr]
+ \frac{m^4\sigma^2}{2}\,\Pi_m^2.
\label{ren-helm}
\eea

In the next section we discuss the determination of the critical temperature.

\section{The critical temperature}
\label{temperature}

The total Helmholtz free energy is obtained by inserting Eq.~(\ref{ren-helm}) in Eq.~(\ref{17}):
\bea
\frac{\overline{\Omega}(\beta, V,\mu, \xi)}{V} &=&  \,\left(m^2 - \mu^2
+ \lambda\xi^2 - m^4\sigma^2\,\Pi_m\right)\xi^2
\nn\\ 
&+& \frac{1}{\beta}\int\frac{d{\bf p}}{(2\pi)^3}\biggr[\ln\left(1-e^{-\beta\Omega_{+}}\right)
\nn\\
&+&\ln\left(1-e^{-\beta\Omega_{-}}\right)\biggr]
- \frac{m^4\sigma^2}{2}\,\Pi_m^2.
\eea
Since we are working in the mean-field approximation, $\lambda\xi \ll 1$, we neglect
contributions coming from terms proportional to $\lambda^2\xi^4$ in the definition of
$\Omega_{\pm}(\p,\xi)$ in Appendix~\ref{app1}. This leads to
\bea
\Omega_{\pm}(\p,\xi) \approx \omega(\p)\pm \mu, \hspace{0.5cm}
\frac{1}{W_{\pm}(\p,\xi)} \approx \frac{1}{\omega(\p)},
\label{aprox}
\eea
with $\omega(\p) = \sqrt{{\p}^2+M^2}$ and $M^2=m^2+4\lambda\xi^2$.  Hence
\bea
&&\frac{\overline{\Omega}(\beta, V,\mu, \xi)}{V} =  \,\left(m^2 - \mu^2
+ \lambda\xi^2 - m^4\sigma^2\,\Pi_m\right)\xi^2
\nn\\ &&
+ \frac{1}{\beta}\int\frac{d{\bf p}}{(2\pi)^3}\biggr[\ln\left(1-e^{-\beta(\omega(\p) + \mu)}\right)
\nonumber\\ &&
 \,+\ln\left(1-e^{-\beta(\omega(\p) - \mu)}\right)\biggr]
- \frac{m^4\sigma^2}{2}\,\Pi_m^2,
\label{ren-therm}
\eea
with
\bea
\Pi_m(\beta, \mu, \xi) &=& \frac{1}{2}\int\frac{d\bf{p}}{(2\pi)^3}
\frac{1}{\omega(\p)}\frac{1}{e^{\beta(\omega(\p) + \mu)}-1}
\nn\\
&+&\frac{1}{2}\int\frac{d\bf{p}}{(2\pi)^3}\frac{1}{\omega(\p)}
\frac{1}{e^{\beta(\omega(\p) - \mu)}-1}.
\label{pim-36}
\eea
From Eq.~(\ref{ren-therm}), one may read off the classical energy density,
i.e. the Helmholtz free energy density at zero temperature:
\beq
\frac{\Omega_{cl}(V, \mu, \xi)}{V} =  (m^2 - \mu_0^2 + \lambda\xi^2)\xi^2,
\label{cla-therm}
\eeq
where $\mu_0 = \mu(T = 0)$ is the chemical potential at zero temperature. To such
a quantity one should add the contributions coming from the zero-point energy of the
fields as well as the divergent vacuum term. This leads us to a divergent vacuum energy
density. Its regularization and renormalization are discussed at length in
Appendix~\ref{app4} and the final result is that the renormalized vacuum
energy density equals the classical contribution, $\Omega_{cl}/V$.

As discussed in Ref.~\cite{kapusta}, the parameter $\xi$ is not determined \textit{a priori}
and it should be treated as a variational parameter, related to the charge carried by the
condensed particles. At fixed $\beta$ and $\mu$, the free energy is an extremum with respect
to variations of $\xi$. The derivative of Eq.~(\ref{ren-therm}) with respect
to $\xi$ implies that $\xi =0$ unless
\beq
\xi_0^2 =  \frac{1}{2 \lambda}\left[\mu^2 - m^2 - 4 \lambda_{\rm eff}(\sigma)
\, \Pi_m(\beta, \mu, \xi_0)\right],
\label{chemical}
\eeq
where
\beq
\lambda_{\rm eff}(\sigma) = \lambda - \frac{\sigma^2 m^4 }{4} .
\label{lambda_eff}
\eeq
Here we have used Eq.~(\ref{aprox}) and neglected second order terms proportional to
$\lambda$ $m^4\sigma^2$ in accord with the assumption of weak disorder and the use of
the mean field approximation.

Since $\xi$ is related to the charge carried by the condensate, at the transition we have
$\xi_0 = 0$ and then
\beq
\mu_c^2 - m^2 - 4\lambda_{\rm eff}(\sigma) \Pi_m(\beta_c, \mu_c, 0) = 0.
\label{tc}
\eeq
This equation gives the critical temperature $T_c = \beta^{-1}_c$ in terms of the critical
chemical potential $\mu_c = \mu(T_c)$ as function of the parameters of the model:
$m$, $\lambda$ and~$\sigma^2$. To~clarify the influence of disorder on $T_c$, let us
consider the behavior of the critical temperature in the ultrarelativistic
limit of Eq.~(\ref{tc}). Since this is akin to performing a high-temperature expansion,
we follow the technique developed in Ref.~\cite{haber,haber2} to obtain an analytical expression
for the critical temperature. The relevant formulae are collected in Appendix~\ref{app5}.

Inserting in Eq.~(\ref{tc}) the expression for the ultrarelativistic limit (i.e., $\beta m \ll 1$)
of $\Pi_m(\beta, \mu, \xi_0)$, Eq.~(\ref{pim58}), one gets for the critical temperature
\bea
T_c^{2} = \left[ 1 + \frac{\sigma^2 m^4}{4\lambda_{\rm eff}(\sigma)}\right]  T^2_0,
\label{Tc}
\eea
where $T_0$ is the mean field critical temperature in the absence of disorder~\cite{kapustax}:
\beq
T^2_0 = \frac{3}{\lambda}(\mu_c^2 - m^2) .
\label{T0}
\eeq
Clearly, disorder implies in an increase of the condensation temperature. Also, there is
a critical value for $\sigma^2$ for which Bose-Einstein condensation occurs only if
$\mu_c = m$  -- namely, $\sigma^2_c = 4 \lambda/m^4$, which implies in
$\lambda_{\rm eff}(\sigma_c) = 0$. The condition $\mu_c = m$ is precisely the one for
condensation of the free relativistic Bose gas~\cite{kapustax}. While one should keep
in mind that there might be important nonperturbative corrections to the precise value
of critical value of the noise intensity, $\sigma_c$, it is clear that noise has induced
an effective negative self-coupling for the scalar field that competes with the original
repulsive coupling $\lambda > 0$.

\begin{figure}[htb]
\begin{center}
\includegraphics[height=80mm,width=80mm]{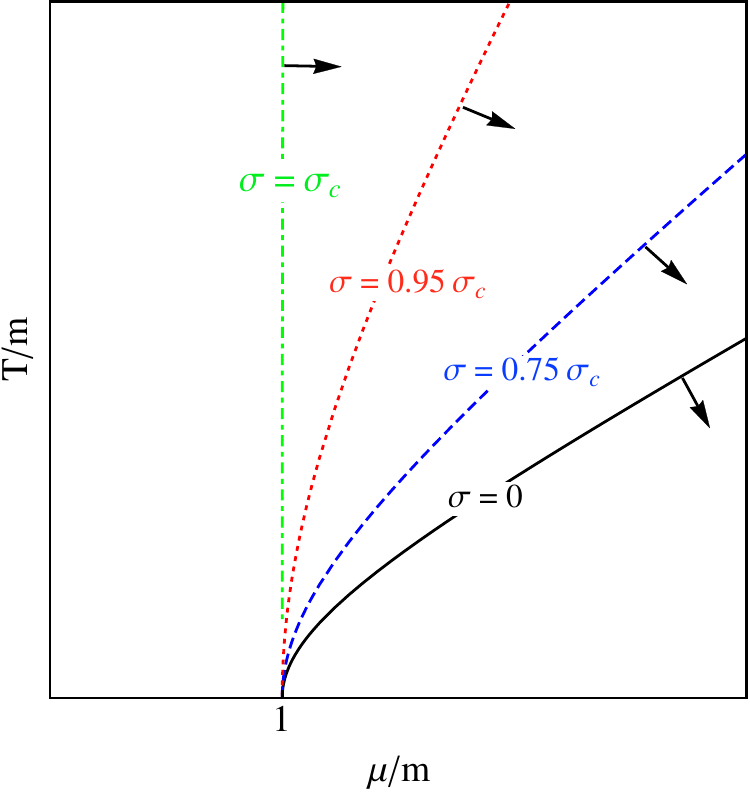}
\caption{Temperature vs chemical potential phase-diagram for $\lambda =0.1$ and four different
values of the noise intensity $\sigma$. The arrows indicate the region of condensation, where
$\xi_0 \neq 0$.}
\label{1}
\end{center}
\end{figure}

The $T$-$\mu$ phase diagram $\lambda = 0.1$ is shown in Fig.~\ref{1}. We use rescaled
quantities $T/m$ and $\mu/m$. The noiseless mean field result is indicated by the (black)
solid curve, which is the standard result~\cite{kapustax}. The vertical (green) dash-dotted
line is for $\sigma = \sigma_c$, for which the effective coupling $\lambda_{\rm eff}(\sigma)$
vanishes and, as said above, condensation occurs for $\mu_c = m$.

The chemical potential is a temperature-dependent parameter related to the total charge.
This is discussed in the next section.

\section{Temperature dependence of the chemical potential}

\subsection{Fixed charged density}

Here we investigate the $T$ dependence of the chemical potential for a fixed total charge
density $\rho = Q/V$, with $Q > 0$, i.e. particles outnumber antiparticles. As usual, the
charge density is calculated by differentiating with respect to $\mu$ the Helmholtz
free energy at its minimum ($\xi=\xi_0$):
\begin{equation}
\rho = -\frac{1}{V}\left(\frac{\partial
\overline{\Omega}}{\partial \mu}\right)_{\beta, V, \xi},
\label{charge}
\end{equation}
where it should be understood that $\overline{\Omega} = \overline{\Omega}(\beta, V, \mu,\xi)$.
For temperatures above the critical temperature, one has $\xi_0 = 0$; below the critical
temperature $\xi_0$ is a solution of Eq.~(\ref{chemical}). Inserting Eq.~(\ref{ren-therm})
in the above Eq.~(\ref{charge}) and employing Eq.~(\ref{aprox}) one obtains for $\rho$:
\beq
\rho = \left(2 \mu+ m^4\sigma^2\frac{\partial \Pi_m}{\partial \mu}\right)\xi_0^2
+ \rho^*(\beta, \mu, \xi_0) + \rho_I(\beta, \mu, \xi_0),
\label{sach}
\eeq
where $\rho^*$ is the mean-field thermal contribution:
\bea
\rho^*(\beta,\mu, \xi) &=& \int\frac{d\p}{(2\pi)^3}\biggl[\frac{1}{e^{\beta(\omega(\p)-\mu)}-1}
\nn\\
&-&\frac{1}{e^{\beta(\omega(\p)+\mu)}-1}\biggr],
\label{rhof}
\eea
and $\rho_I$ is the contribution due to disorder
\beq
\rho_I(\beta, \mu, \xi)\, = \sigma^2 m^4 \, \Pi_m \frac{\partial \Pi_m}{\partial \mu}.
\label{rhoi}
\eeq

For a fixed $\rho$, Eq.~(\ref{sach}) can be formally inverted to give the chemical potential
as a function of the temperature. Using the expressions derived in Appendix~\ref{app5}, one
obtains for $\rho^*$:
\beq
\rho^*(\beta, \mu, \xi) \approx \frac{\mu}{3 \beta^2} + \frac{M^2\mu}{4 \pi^2}
- \frac{\mu^3}{6\pi^2},
\label{rho-ultra}
\eeq
and for $\rho_I$:
\beq
\rho_I(\beta, \mu, \xi) \approx \frac{\sigma^2m^4 \mu}{16\pi^2} \left(\frac{\mu}{2\pi^2} - \frac{1}{3\beta^2}\right).
\label{rhoi-ultra}
\eeq
Inserting these results in Eq.~(\ref{sach}), one obtains
\beq
\rho \approx 2 \mu\left[1 + \frac{1}{2\pi^2} \lambda_{\rm eff}(\sigma) \right]\xi_0^2
+\frac{\mu}{3 \beta^2}\left(1 - \frac{\sigma^2m^4}{16\pi^2}\right).
\label{sach-ultra}
\eeq
The first term is the charge density associated with the condensate (zero-momentum mode)
\beq
\rho_c \approx 2 \mu \left[1 + \frac{1}{2\pi^2} \lambda_{\rm eff}(\sigma) \right] \xi_0^2,
\label{charge-con}
\eeq
and the second is the charge density associated with the thermal particle excitations (finite-momentum modes)
\beq
\rho_{th} \approx \frac{\mu}{3 \beta^2}\left(1 - \frac{\sigma^2m^4}{16\pi^2}\right).
\label{charge-therm}
\eeq
Using Eqs.~(\ref{chemical}), (\ref{tc}), and~(\ref{pim58}), one obtains for the condensate $\xi_0$:
\beq
\xi_0^2 \approx \frac{1}{2\lambda}\left[\mu^2 - \mu_c^2 - \frac{\lambda_{\rm eff}(\sigma)}{3}\left(\frac{1}{\beta^{2}}
- \frac{1}{\beta_c^{2}}\right)\right].
\label{chemical-ultra}
\eeq
At the critical temperature $\xi_0 = 0$ and
\beq
\rho \approx \frac{\mu_c }{3\beta_c^2}\left(1 - \frac{\sigma^2m^4}{16\pi^2}\right).
\label{rho-critical}
\eeq

Finally, one can obtain the expression of the chemical potential in function of the the
temperature. Inserting Eq.~(\ref{chemical-ultra}) in Eq.~(\ref{sach-ultra}), one obtains
a cubic equation for $\mu = \mu_\sigma(T)$ in terms of the total charge density $\rho$. For temperatures just below
$T_c$ the approximate solution is given by
\bea
\mu_\sigma(T) &\approx& \mu_c + \frac{\lambda\left(T_c^2 - T^2\right)}{6\tilde{\mu}_c^2 + \lambda T_c^2}\,\mu_c
\nn\\
&\times&\left[\left(1 + \frac{3\lambda}{4\pi^2}\right) \frac{\sigma^2m^4}{4\lambda} - \frac{\lambda}{2\pi^2}\right],
\label{chemical-critical}
\eea
where
\beq
\tilde{\mu}_c^2 = \left[1 + \frac{1}{2\pi^2} \lambda_{\rm eff}(\sigma) \right] \, \mu_c^2 .
\eeq

A close inspection of Eq.~(\ref{chemical-critical}) reveals that for $\sigma = 0$ one has the mean-field
solution. As the temperature is reduced beyond $T_c$, the mean-field $\mu(T)$ continues to decrease,
even though for sufficiently low temperatures such an expression ceases to be a good approximation.
This is in agreement with the usual results of Refs.~\cite{kapustax,haber,haber2}. This scenario is modified
for $\sigma \neq 0$. Neglecting the term $3\lambda/4\pi^2$, in order to keep the same behavior
one must require that $m^4\sigma^2 < 2\lambda^2/\pi^2$. This situation respects the stability assumption:
$m^4\sigma^2 \ll \lambda$. However the case in which $m^4\sigma^2 > 2\lambda^2/\pi^2$ is also possible
provided that the stability condition remains valid. Actually, for the special case
$m^4\sigma^2 \approx 2\lambda^2/\pi^2$, $\mu(T) \approx \mu_c$, even though the system is not at
the critical point. Within the scenario in which $m^4\sigma^2 > 2\lambda^2/\pi^2$, $\mu(T)$ increases
as the temperature is reduced. We interpret this as an energetically non-favorable situation and we
conjecture that disorder may destabilize the condensate. In order to confirm such a conjecture,
one should consider field self-interactions beyond the mean-field approximation employed here, which
is outside the scope of the present work.

\begin{figure}[htb]
\begin{center}
\includegraphics[height=80mm,width=80mm]{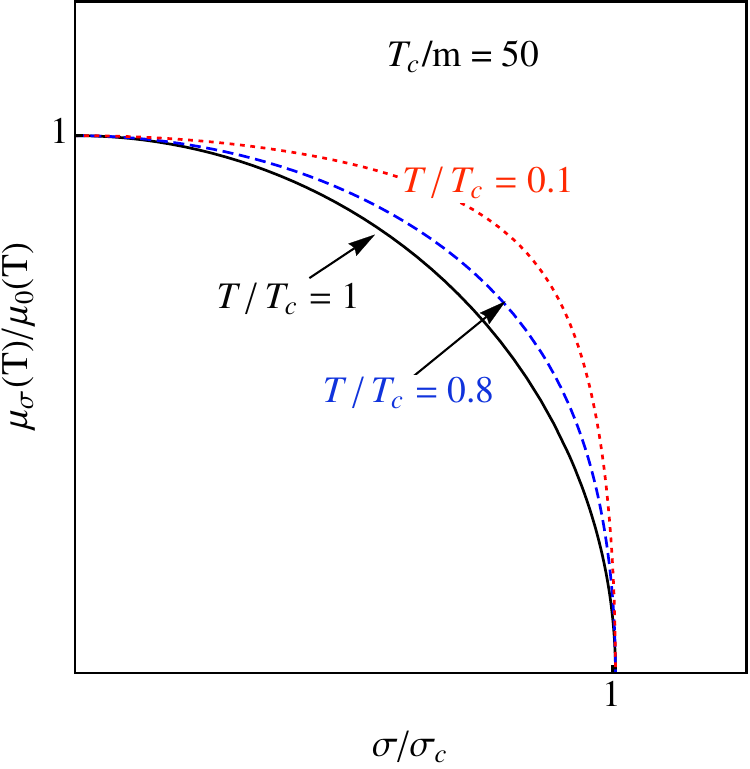}
\caption{Chemical potential as a function of $\sigma$ for three different temperatures $T \leq T_c$,
for $\lambda = 0.1$ and $T_c/m = 50$.}
\label{2}
\end{center}
\end{figure}

Let us analyze the behavior of $\mu_\sigma(T)$, Eq.~(\ref{chemical-critical}), as a function of the noise
intensity $\sigma$ and for a fixed temperature $T \leq T_c$, depicted in Fig.~\ref{2}. For small values of
$\sigma$, the chemical potential is approximately constant and thereafter it starts to decrease. Close to
the critical value of $\sigma$, the chemical potential is close to zero: this is the region where the
induced interactions balance the field self-interactions; for $\sigma > \sigma_c$, the mean-field
solution is destabilized.

Substituting Eq.~(\ref{chemical-critical}) in~(\ref{charge-con}) and using~(\ref{rho-critical})
one gets, for $0 \ll T \lesssim T_c$ and $\lambda, m^4\sigma^2 \ll \beta m \ll 1$
\beq
\rho_c \approx \rho\left[1 - \left(\frac{\beta_c}{\beta}\right)^2\right].
\label{charge-conultra}
\eeq
This is the same behavior as found for the ground-state charge density of the ideal gas,
see e.g. Ref.~\cite{haber,haber2}. This corresponds to a temperature-dependent $\xi_0$ given by
\beq
\xi_0^2 \approx \frac{1}{6} \left(T_c^2 - T^2\right)\,
\left[\frac{1- \sigma^2 m^4/16\pi^2}{1+\lambda_{\rm eff}(\sigma)/2\pi^2}\right] .
\eeq

For completeness, let us present the critical temperature as a function of the fixed charge density $\rho$:
\begin{equation}
\rho \approx \frac{m}{3\beta_c^2}\left[1+\frac{\lambda}{6(\beta_c m)^2}
- \frac{\sigma^2 m^4}{24(\beta_c m)^2}\right].
\label{rhofix}
\end{equation}
Hence, the ultrarelativistic critical temperature in the weak-disorder limit is given by:
\beq
\beta_c^{-1} \approx \left(\frac{3\rho}{m}\right)^{1/2}\left[1 - \frac{\lambda}{12(\beta_c m)^2}
+ \frac{\sigma^2 m^4}{48(\beta_c m)^2}\right].
\label{tcriticalultra}
\eeq
Solving this equation by iteration one arrives at a power series expansion of $\beta_c^{-1}$
in the effective coupling $m^4\sigma^2$. At first order, one has
\beq
\beta_c^{-1} \approx \beta_u^{-1}\left[1 - \frac{\lambda_{\rm eff}(\sigma)}{12(\beta_u m)^2}\right],
\label{tcriticalultra1}
\eeq
where $\beta_u^{-1}$ is the ultrarelativistic critical temperature for the free gas
\beq
\beta_u^{-1} = \left(\frac{3\rho}{m}\right)^{1/2}.
\eeq
As above, we find a positive shift in the critical temperature due to random fluctuations.
Again, for $\lambda_{\rm eff}(\sigma_c) = 0$ the transition temperature is the same as the free case,
i.e. the system behaves effectively as a free Bose gas.

\subsection{Charge and entropy fixed}

As discussed in the Introduction, relativistic Bose-Einstein condensation has important cosmological
implications. In such a context, in most cases the volume $V$ changes with temperature but the net
total charge $Q$ and the entropy ${\cal S}$ remain constant. Therefore it is crucial to study the
temperature evolution of the chemical potential with $Q$ and ${\cal S}$ fixed (as in the early Universe).
The entropy is given by
\beq
{\cal S} = \beta^2\left(\frac{\partial \overline{\Omega}}{\partial \beta}\right)_{V, \mu, \xi},
\eeq
where again it is to be understood that one must set $\xi=\xi_0$ after taking the above derivative.
Inserting Eq.~(\ref{ren-therm}) in the above expression and employing Eq.~(\ref{aprox}), one gets
\beq
\frac{{\cal S}}{V} = \frac{{\cal S}_{MF}}{V} + \frac{{\cal S}_I}{V},
\label{entropy}
\eeq
where the mean-field entropy is given by
\beq
\frac{{\cal S}_{MF}}{V} = - I(\beta, \mu, \xi_0)+ \beta\,\frac{\partial I(\beta, \mu, \xi_0)}{\partial \beta},
\eeq
with
\bea
I(\beta, \mu, \xi) &=& \int\frac{d{\bf p}}{(2\pi)^3}\biggr[\ln\left(1-e^{-\beta(\omega(\p) + \mu)}\right)
\nn\\
&+&\ln\left(1-e^{-\beta(\omega(\p) - \mu)}\right)\biggr],
\eea
whereas the corrections due to random fluctuations are given by
\beq
\frac{{\cal S}_I}{V} = - \sigma^2 m^4 \,\beta^2\left(\xi_0^2 + \Pi_m\right)
\frac{\partial \Pi_m}{\partial \beta} ,
\eeq
with $\Pi_m = \Pi_m(\beta, \mu, \xi_0)$ and $\omega(\p) = \sqrt{\p^2 + m^2 + 4\lambda\xi_0^2}$.
Using the expressions derived in Appendix~\ref{app5}, one obtains
\beq
\frac{{\cal S}}{V} \approx \frac{4\pi^2}{45\beta^3}
+ \left(\xi_0^2 + \frac{1}{12\beta^2}\right)\frac{\sigma^2 m^4}{6\beta}
\eeq
Taking the ratios of the associated charge densities $\rho_c$ and $\rho_{th}$, given
in Eqs.~(\ref{charge-con}) and~(\ref{charge-therm}), respectively, with the
expression above leads to
\beq
\frac{Q_c}{{\cal S}} \approx \frac{45\mu\beta^3}{2\pi^2}
\left(1 - \frac{5\sigma^2 m^4}{32\pi^2}\right)\tilde{\xi}_0^2
\label{netc}
\eeq
and
\beq
\frac{Q_t}{{\cal S}} \approx \frac{15\mu\beta}{4\pi^2}
\left(1 - \frac{7 \sigma^2 m^4}{32\pi^2}\right),
\label{net}
\eeq
where terms proportional to $\beta^2 \sigma^2 m^4$ were dropped and
\beq
\tilde{\xi}_0^2 = \xi_0^2 \left[1 + \frac{1}{2\pi^2}\lambda_{\rm eff}(\sigma)\right],
\eeq
with $\xi_0^2$ given by Eq.~(\ref{chemical-ultra}). The sum of Eqs.~(\ref{netc}) and~(\ref{net})
produces a term independent of $T$. In the high-temperature region, the total net charge $Q$ is given by
Eq.~(\ref{net}). Lowering the temperature, there would be a point such that
\beq
\frac{Q}{{\cal S}} \approx \frac{15\hat{\mu}_c\hat{\beta}_c}{4\pi^2}
\left(1 - \frac{7 \sigma^2 m^4}{32\pi^2}\right),
\label{net-total}
\eeq
where $1/\hat{\beta}_c = \hat{T}_c$ is the critical temperature for a fixed $Q/{\cal S}$ and $\hat{\mu}_c$
is the associated critical chemical potential. After a little algebra one can show that
\bea
\hat{\beta}_c^2 &\approx& \frac{1}{m^2}\biggl[\frac{\sigma^2 m^4}{12} - \frac{\lambda}{3}
\nn\\
&+&\frac{16\pi^4}{225}\left(1 + \frac{7 \sigma^2 m^4}{16\pi^2}\right)\left(\frac{Q}{{\cal S}}\right)^2\biggr].
\eea
Remembering the stability assumption mentioned earlier and assuming that $Q/{\cal S}$ is sufficiently small,
the above expression shows that $\hat{\beta}_c^2 < 0$ and thus in this case adiabatic cooling will not lead
to symmetry breaking. Thence all the net charge is correctly given by Eq.~(\ref{net}) and there will be no
Bose-Einstein condensation. This result is similar to the one in the absence of random fluctuations~\cite{haber,haber2}.

Thus, there appears that assuming that $Q/{\cal S}$ is large enough so that $\hat{\beta}_c^2 > 0$, one could
expect Bose-Einstein condensation to take place. However, this expectation falls apart when one soon realizes
that there are other inconsistencies plaguing this particular case. Summing Eqs.~(\ref{netc}) and~(\ref{net})
and inserting Eq.~(\ref{chemical-ultra}) in the result yields a cubic equation for
$\mu(T)$ in terms of the temperature and $Q/{\cal S}$. If $\hat{\mu}_c$ is the value of the chemical
potential at the critical temperature, then just below $\hat{T}_c$ the approximate solution is
\bea
\mu &\approx& \hat{\mu}_c + \frac{2\lambda\left(\hat{T}_c^2 - T^2\right)}{6\tilde{\tilde{\mu}}_c^2
+ \lambda \hat{T}_c^2}\,\hat{\mu}_c
\nn\\
&\times&\left[\frac{\sigma^2 m^4}{8\lambda}\left(1 + \frac{33\lambda}{16\pi^2}\right) - \frac{3}{4}
-\frac{\lambda}{4\pi^2}\right],
\label{chemical-critical2}
\eea
where again terms proportional to $\lambda \sigma^2 m^4$ and $\sigma^4 m^8$ were dropped and
\beq
\tilde{\tilde{\mu}}_c^2 = \hat{\mu}_c^2\left(1 + \frac{\lambda}{2\pi^2}
- \frac{9 \sigma^2 m^4}{32\pi^2}\right).
\eeq
Inserting Eq.~(\ref{chemical-critical2}) in Eq.~(\ref{netc}) one gets
\beq
\frac{Q_c}{{\cal S}} \approx - \frac{Q}{2{\cal S}}\left[1-\left(\frac{\hat{\beta}_c}{\beta}\right)^2\right],
\eeq
for $0 \ll T \lesssim T_c$ and $\lambda, \sigma^2 m^4 \ll \beta m \ll 1$, where $Q/{\cal S}$ is given by Eq.~(\ref{net-total}).
In the present context, this corresponds to the following temperature-dependent $\xi_0$
\beq
\xi_0^2 \approx - \frac{1}{12}(\hat{T}_c^2 - T^2) \left[ 1 - \frac{\lambda}{2\pi^2} + \frac{\sigma^2 m^4}{16\pi^2}\right].
\eeq
Note that $\xi_0^2 < 0$, which is clearly unphysical; on the other hand $Q_c/{\cal S}$ is also negative
which contradicts our initial assumption that particles outnumber antiparticles. Thence for a fixed $Q/{\cal S}$
there will be no Bose-Einstein condensation, a result already expected within the mean-field theory in the absence
of disorder. As emphasized in Ref.~\cite{haber,haber2} this is due to the fact that $m^2 > 0$. We conclude that a nonstatic
weak disorder does not change such a behavior.

\section{Conclusions and Perspectives}
\label{conclude}

In this paper we investigated the effect of weak disorder on a weakly interacting relativistic charged
scalar field in thermal equilibrium with a reservoir. We studied the effect of coupling
of a random field to the scalar field in the situation where Bose-Einstein condensation
takes place. We considered a quenched disorder which couples linearly to the
mass of the scalar field, just as in the random-temperature Landau-Ginzburg model.
After performing noise averages of the free energy, we obtained the corrections to
the mean field critical temperature for the interacting Bose gas at finite density.

We have shown that the effect of the randomness is to increase the critical temperature for fixed
charge density $\rho = Q/V$. We observed significant differences from the mean-field
temperature dependence of the chemical potential as the strength of the
noise intensity increases. In particular, we found that for a critical noise intensity,
the model behaves as a free field theory. In addition, having in mind application in the
physics of the early Universe, we have investigated the temperature dependence of the chemical
potential with fixed total charge and entropy. We found that there is no Bose-Einstein condensation
for a fixed charge to entropy ratio in the presence of weak disorder.

Naturally, one should keep in mind that this is a result valid for weak disorder and obtained in the
framework of a perturbative expansion in the noise intensity. It remains to be seen if the same result is
attainable with a nonperturbative  calculation, e.g. using a replica trick. For $\tau$-independent noise,
application of the replica-trick consists in the following~\cite{dot}: using the fact that one can write
$\ln Z= \lim_{n\rightarrow 0} {(Z^n-1)}/{n}$, one has that $\overline{\ln Z[\nu]}
= \lim_{n\rightarrow 0}{(Z_{n}-1)}/{n}$, where $Z_{n}=\overline{Z^{\,n}[\nu]}$; the $Z_{n}$'s
are interpreted as the partition functions of new systems, formed from $n$ statistically independent
copies of the original system. The quenched free energy functional is defined as $F_{q}(h)\equiv - \lim_{n\rightarrow\,0} \,(Z_{n}-1)/{n}$, showing that the quenched free energy
functional can be calculated from a zero-component field theory.

We remark on an important point with respect to the fact that random mass models generate effective
interactions that mimic a negative coupling constant. Many authors claim that non-relativistic bosons only
make sense in a random potential when they present repulsive interactions~\cite{fisher}. Nevertheless, there
are many examples that even for a free theory one can define the theory in a controllable fashion.
For instance, relativistic scalar field models with negative coupling constant were investigated in the
literature and meaningful results were obtained -- see for example Refs.~\cite{prova,brandt1,brandt2,riva,gaw,lang}.
Based on the results obtained in Ref.~\cite{parisi}, where it has been shown that the theory with a negative
coupling constant develops a condensate, Arias~\textit{et al}~\cite{arias} discussed the thermodynamics of a
asymptotically free Euclidean self-interacting scalar field defined in a compact spatial region
without boundaries.

To conclude, we mention that  effects of randomness over quantum fields have been discussed in different physical scenarios. In particular, on the basis of the results of Refs.~\cite{pe9,pe10,pe10a}, it was proposed in a condensed-matter-physics setting an analog model for fluctuations of the light cone~\cite{krein1}. Also, a free massive scalar field in inhomogeneous random media was studied in~\cite{new}. After performing the averages over the random functions, the two- and four-point causal Green's function of the model were presented up to one-loop approximation. Likewise, Refs.~\cite{detector} and~\cite{detector2} investigated
the influence of fluctuations of the event horizon on the transition rate of a two-level system which interact
with a quantum field. More recently studies of effects of light-cone fluctuations over the renormalized vacuum
expectation value of the stress-energy tensor of a real massless scalar field were carried out in Ref.~\cite{tuv}.
In this case the field was defined in a flat space-time with non-trivial topology. In Ref.~\cite{tuv2} the influence
of such random fluctuations upon the zero-point energy associated with a free massless scalar in the presence of
boundaries was investigated. Nonperturbative extensions of such works are under investigation by the authors.

\section*{Acknowledgments}
We thank C. Bessa for helpful discussions. Work partially supported by Conselho Nacional de Desenvolvimento
Cient\'{\i}fico e Tecnol\'ogico -- CNPq, Grants No. 305894/2009-9 and No. 303629/2011-8, and Funda\c{c}\~ao
de Amparo \`a Pesquisa do Estado de S\~ao Paulo -- FAPESP, Grant No. 2013/01907-0, and Funda\c{c}\~ao de Amparo
\`a Pesquisa do Estado do Rio de Janeiro - FAPERJ.

\appendix

\section{Calculation of the partition function}
\label{app1}

In this Appendix we calculate the logarithm of the free partition function given by Eq.~(\ref{z0}). We start by introducing Fourier series to the fields $\chi_1$ and $\chi_2$:
\begin{equation}
\chi_i(\tau, \x) =\biggl(\frac{\beta}{V}\biggr)^{1/2}\sum_n\,\sum_{\p} e^{i(\p\cdot\x + \omega_n\tau)}\chi_{i;\,n}(\p),
\label{fourier}
\end{equation}
with $i = 1,2$ and $\beta\omega_n = 2\pi n$ due to the constraint of periodicity $\chi_i(0, \x) = \chi_i(\beta, \x)$
for all $\x$. Inserting this last result in the free field action given by Eq.~(\ref{s0}) we obtain, after performing an integration by parts:
\begin{equation}
S_0= - \frac{1}{2}\sum_{n,\bf p}
\bigl(\begin{array}{cc}
\chi_{1;-n}(-\p)  & \chi_{2;-n}(-\p)
\end{array}\bigr)
\Theta
\Biggl(\begin{array}{c}
\chi_{1;n}({\p}) \\
\chi_{2;n}(\p)
 \end{array}\Biggr),
\label{s01}
\end{equation}
where we discarded a total derivative term and we defined the matrix $\Theta = \Theta_n(\p)$ as
\begin{equation}
\Theta_n(\p) = \beta^2\Biggl(\begin{array}{cc}
\omega_n^2+\omega_1^2(\p)  - \mu^2& -2\mu\omega_n\\
2\mu\omega_n & \omega_n^2+\omega_2^2(\p) - \mu^2
\end{array}\Biggr),
\label{d}
\end{equation}
with $\omega_i(\p) =\sqrt{\p^2 + m_i^2}$, $i = 1,2$, $m_1^2 = 6\lambda\xi^2 + m^2$ and $m_2^2 = 2\lambda\xi^2 + m^2$. Thus, using Eq.~(\ref{s01}) the logarithm of the free partition function now becomes
\beq
\ln Z_0 = \ln(N(\beta))^2 + \ln J(\beta,\mu)
\label{a4}
\eeq
where
\begin{widetext}
\bea
J(\beta,\mu) &=& \prod_{n}\prod_{\p}\int d\chi_{2;n}({\bf p})\,\exp{\biggl\{-\frac{1}{2}\Bigl[\Theta_n(\p)\Bigr]_{22}\chi_{2;n}({\p})\chi_{2;-n}({-\p})\biggr\}}
\nonumber\\&&
\times\,\int d\chi_{1;n}({\bf p})\,\exp{\biggl\{-\frac{1}{2}\Bigl[\Theta_n(\p)\Bigr]_{11}\chi_{1;n}({\p})\chi_{1;-n}({-\p})\biggr\}}
\nonumber\\&&
\times\,\exp{\biggl\{-\frac{1}{2}\Bigl[\Theta_n(\p)\Bigr]_{12}\chi_{2;n}({\p})\chi_{1;-n}({-\p}) -\frac{1}{2}\Bigl[\Theta_n(\p)\Bigr]_{21}\chi_{1;n}({\p})\chi_{2;-n}({-\p})\biggr\}}.
\label{ai}
\eea
\end{widetext}
Noting that $\chi_{i;-n}(-\p) = \chi_{i}^{*}(\p)$, $i =1, 2$, as required by the reality of the fields $\chi_i(\tau, \x)$, the above integrals are just generic Gaussian integrals. Therefore
\beq
J(\beta,\mu) = \prod_{n}\prod_{\p}\bigl[\det\Theta_n(\p)\bigr]^{-1/2},
\eeq
where we have discarded an overall constant multiplicative factor. Inserting this last expression in Eq.~(\ref{a4}), we get:
\beq
\ln Z_0 = \ln(N(\beta))^2 -\frac{1}{2}\ln\left[\prod_n \prod_{\p}\det\Theta_n(\p)\right],
\label{a7}
\eeq
where
\begin{widetext}
\beq
\ln\left[\prod_n \prod_{\p}\det\Theta_n(\p)\right] = \ln\Biggl\{\prod_n \prod_{\p}\beta^4\biggl[\left(\omega_n^2+\omega_1^2(\p)  
- \mu^2\right)\left(\omega_n^2+\omega_2^2(\p)  - \mu^2\right)
+ 4\mu^2\omega_n^2\biggr]\Biggr\},
\label{fac}
\eeq
\end{widetext}
and, according to Ref.~\cite{bernard}, in the large-volume limit
\beq
\ln(N(\beta)) = -V\ln\beta\sum_{n}\int\frac{d\p}{(2\pi)^3}.
\label{berna}
\eeq
It is possible to factorize the quantity inside the square brackets in Eq.~(\ref{fac}) by defining the ``effective mass"
\beq
 M^2=\frac{m_1^2+m_2^2}{2}+\frac{\lambda^2\xi^4}{\mu^2}.
\label{m-eff}
\eeq
One gets
\bea
&&\left[\omega_n^2+\omega_1^2(\p)  - \mu^2\right]\left[\omega_n^2+\omega_2^2(\p)  - \mu^2\right]
+ 4\mu^2\omega_n^2
\nn\\
&& = \left[\omega_n^2 + \Omega_{+}^2(\p,\xi)\right]\left[\omega_n^2 + \Omega_{-}^2(\p,\xi)\right],
\eea
where
\beq
 \Omega_{\pm}^2(\p,\xi)=\left(\omega(\p)\pm \mu\right)^2-\frac{\lambda^2\xi^4}{\mu^2},
\label{o-eff}
\eeq
with $\omega(\p) = \sqrt{{\p}^2+M^2}$. Hence
\bea
&&\ln Z_0 =  -\frac{1}{2}\sum_{n,\p}\ln\left\{\beta^2\bigl[\left(\omega_n^2 + \Omega_{+}^2\right)\bigr]\right\}
\nn\\ &&
\,-\frac{1}{2}\sum_{n,\p}\ln\left\{\beta^2\bigl[\left(\omega_n^2 + \Omega_{-}^2\right)\bigr]\right\}.
\nn\\ &&
\,-2 V\ln\beta\sum_{n}\int\frac{d\p}{(2\pi)^3},
\label{i4}
\eea
where $\Omega_{\pm}= \Omega_{\pm}(\p,\xi)$. The frequency sums can be performed using standard procedures~\cite{bellac} 
and the result is given by Eq.~(\ref{a10}).

In particular, since for a given $n$ and $\p$ the propagators can be expressed as functional derivatives of the partition 
function~\cite{kapusta}
\beq
{\cal D}_{ij}(\omega_n,\p) = 2 \left({\cal D}^0_{ij}\right)^2\frac{\delta\ln Z[\nu]}{\delta {\cal D}^0_{ij}}.
\label{3}
\eeq
$i, j = 1, 2$, one notes that the zero-order propagators ${\cal D}^0_{ij}(\omega_n,\p)$ are given by
\beq
{\cal D}^0_{ij}(\omega_n,\p) = -2\beta^2\,\frac{\delta\ln Z_0}{\delta [\Theta_n(\p)]_{ij}}.
\label{b4}
\eeq
%

\section{Renormalization of propagators}
\label{app3}

Here we examine the renormalization of the finite-temperature propagators ${\cal D}^0_{11}(\omega_n,\p)$ and ${\cal D}^0_{22}(\omega_n,\p)$. 
Following Ref.~\cite{kapusta} we define the self-energy $\Pi_1= \Pi_1(\omega_n,\p)$ with respect to the averaged propagator 
$\overline{{\cal D}_{11}}$ as
\beq
\overline{{\cal D}_{11}(\omega_n,\p)} = \left(1 + {\cal D}^0_{11}\,\Pi_1\right)^{-1}{\cal D}^0_{11}.
\label{4}
\eeq
A similar expression holds for $\Pi_2= \Pi_2(\omega_n,\p)$ which is the self-energy with respect to the averaged propagator 
$\overline{{\cal D}_{22}}$. Hence, recalling Eqs.~(\ref{3}) and~(\ref{b4}), we get, up to second order in $\nu$:
\beq
\left(1 + {\cal D}^0_{11}\,\Pi_1\right)^{-1} = 1 + 2{\cal D}^0_{11}\,\frac{\delta\overline{\ln Z_I[\nu]}}{\delta {\cal D}^0_{11}},
\label{6}
\eeq
and
\beq
\left(1 + {\cal D}^0_{22}\,\Pi_2\right)^{-1} = 1 + 2{\cal D}^0_{22}\,\frac{\delta\overline{\ln Z_I[\nu]}}{\delta {\cal D}^0_{22}}.
\label{6a}
\eeq
Therefore, inserting Eq.~(\ref{zi-ave1}) in the above equations and expanding their left-hand sides to first order yields the following expressions for the self-energies
\bea
\Pi_1(\omega_n,\p) &&= -2 m^4\sigma^2\xi^2
\nn\\ &&
-\frac{m^4\sigma^2}{\beta V}\sum_{n,\p}\,{\cal D}^0_{11}(\omega_n,\p),
\label{10}
\eea
and
\beq
\Pi_2(\omega_n,\p) = - \frac{m^4\sigma^2}{\beta V}\sum_{n,\p}\,{\cal D}^0_{22}(\omega_n,\p).
\label{11}
\eeq
On the other hand, remembering Eq.~(\ref{b4}) one gets
\begin{widetext}
\beq
\frac{1}{\beta V}\sum_{n,\p}\,{\cal D}^0_{ii}(\omega_n,\p)
= \frac{1}{4}\int\frac{d\bf{p}}{(2\pi)^3}\frac{1}{W_{-}(\p,\xi)}\biggl[1+\frac{2}{e^{\beta\Omega_{-}}-1} \biggr] 
+ \frac{1}{4}\int\frac{d\bf{p}}{(2\pi)^3}\frac{1}{W_{+}(\p,\xi)}\biggl[1+\frac{2}{e^{\beta\Omega_{+}}-1} \biggr],
\label{12}
\eeq
\end{widetext}
for $i=1,\,2$ and
$$
\frac{1}{W_{\pm}(\p,\xi)} = \frac{1}{\Omega_{\pm}(\p,\xi)} \pm \frac{\mu}{\omega\Omega_{\pm}(\p,\xi)}.
$$
In Eq.~(\ref{12}) we consider $V$ to be large compared to all other physical lengths so we can replace the sum over 
$\p$ with an integral. In this way we have
\bea
&\Pi_1(\omega_n,\p)&\, = - m^4\sigma^2\biggl[2\xi^2 +\Pi_{v+} + \Pi_{v-}
\nn\\
&&+\,\Pi_{m+}(\beta,\xi) + \Pi_{m-}(\beta,\xi)\biggr],
\label{self1}
\eea
and
\bea
&\Pi_2(\omega_n,\p)&\, = - m^4\sigma^2\biggl[\Pi_{v+} + \Pi_{v-}
\nn\\
&&+\,\Pi_{m+}(\beta,\xi) + \Pi_{m-}(\beta,\xi)\biggr],
\label{self2}
\eea
where we have defined
\beq
\Pi_{v\pm} =  \frac{1}{4}\int\frac{d\bf{p}}{(2\pi)^3}\frac{1}{W_{\pm}(\p,\xi)},
\eeq
and
\beq
\Pi_{m\pm}(\beta,\xi) = \frac{1}{2}\int\frac{d\bf{p}}{(2\pi)^3}\frac{1}{W_{\pm}(\p,\xi)}\frac{1}{e^{\beta\Omega_{\pm}}-1}.
\label{pim}
\eeq
Since $\Pi_v$ is a divergent quantity, in order to avoid physically meaningless results the following counterterm must be 
added to the original action:
\bea
\delta{S} &&= -\delta \sigma^2\int_{0}^{\beta}d\tau\int_{V}d\x\,\varphi\varphi^*
\nn\\ &&
= -\frac{\delta \sigma^2}{2}\int_{0}^{\beta}d\tau\int_{V}d\x\,\Bigl(2\xi^2 + \chi_1^2 + \chi_2^2\Bigr),
\eea
where we have droped terms linear in $\chi_1$ and $\chi_2$. Treating this as an additional interaction, we see from Eq.~(\ref{zi}) 
that to lowest order this counterterm contributes to $\ln Z_I$ as
\beq
-(\beta V)\delta \sigma^2\xi^2 -\frac{\delta \sigma^2}{2}\int_{0}^{\beta}d\tau\int_{V}d\x\,\Bigl(\bigl\langle \chi_1^2  \bigr\rangle
+ \bigl\langle \chi_2^2 \bigr\rangle\Bigr).
\eeq
The counterterm should be chosen so that
\beq
\delta \sigma^2 - m^4\sigma^2(\Pi_{v-} + \Pi_{v+}) = 0.
\eeq
In this way we get a finite result for the propagators. Whence, collecting the above results, the contribution to $\ln Z$ up to second 
order in the noise will be
\bea
\overline{\ln Z_I[\nu]} = (\beta V)\frac{m^4\sigma^2}{2}\Bigl[\Pi_m(2\xi^2 + \Pi_m) - \Pi_v^2\Bigr],
\label{zeta-zi}
\eea
where one has that $\Pi_m = \Pi_{m+}(\beta,\xi) + \Pi_{m-}(\beta,\mu,\xi)$ and also $\Pi_{v} = \Pi_{v+} + \Pi_{v-}$.

\section{Renormalization of the vacuum energy density}
\label{app4}

In this Appendix we discuss the renormalization of the classical energy density, Eq.~(\ref{cla-therm}). In the expression~(\ref{ren-therm}) 
we have neglected the shift in $\Omega_{cl}(\xi)$ coming from the zero-point energy density of the vacuum as well as the divergent 
contribution $\Pi_{v}$ which results from the renormalization of the propagators considered in detail in the previous Appendix 
[e.g., see Eq.~(\ref{zeta-zi})]. Since we are in the mean field approximation, $\lambda\xi \ll 1$, we take into account the same 
approximation mentioned in Sec.~\ref{temperature}. Namely, we neglect the contributions coming from the terms proportional to 
$\lambda^2\xi^4$ in the definition of $\Omega_{\pm}(\p,\xi)$ in Eq.~(\ref{o-eff}). This means that, in this approximation the 
zero-point energy density is given by:
$$
E_0 = \int\frac{d{\p}}{(2\pi)^3}\left(\frac{\Omega_{+}}{2} + \frac{\Omega_{-}}{2}\right) = \int\frac{d{\bf p}}{(2\pi)^3}\,\omega(\p),
$$
whereas  $\Pi_{v}$ becomes
$$
\Pi_{v} = \Pi_{v+} + \Pi_{v-} =  \frac{1}{2}\int\frac{d\bf{p}}{(2\pi)^3}\,\frac{1}{\omega(\p)}.
$$
As a regularization procedure we simply choose to place a high-momentum cutoff $\Lambda_c$ on the integration over $|\p|$. In this way, we get
\beq
 E_0 = \frac{1}{64\pi^2}\left[4M^2\Lambda_c^2-2M^4\ln\left(\frac{\Lambda_c^2}{M^2}\right)-M^4\right],
\label{zero-energy}
\eeq
and
\beq
\Pi_{v} = \frac{1}{32 \pi^2}\left[\Lambda_c^2 - M^2\ln\left(\frac{\Lambda_c^2}{M^2}\right)\right],
\label{div}
\eeq
where, due to the aproximation earlier observed, $M^2=m^2+4\lambda\xi^2$. Also, in Eqs.~(\ref{zero-energy}) 
and~(\ref{div}) we have dropped constants and terms which vanish as $\Lambda_c\to\infty$. In order to renormalize 
the vacuum energy density, we demand that the final result should be independent of $\Lambda_c$. Also, we require 
its minimum to be at the same location as the classical energy density, i.e., at $\xi_c^2 = (\mu_0^2 - m^2)/2\lambda$. 
This is achieved by adding to the original action counterterms which depend on the bare parameters $m^2$ and $\lambda$ 
as well as on $\Lambda_c$. In addition, one should specify a suitable set of normalization conditions. Here we choose
\bea
&&\frac{d^2 \overline{\Omega(0, \mu_0, \xi)}}{d\xi^2}|_{\xi = \xi_c} = 4 \left (\mu_0^2 - m^2\right)
\nn\\[0.3true cm]
&&\frac{d^4 \overline{\Omega(0, \mu_0, \xi)}}{d\xi^4}|_{\xi = \xi_c} = 24 \lambda.
\label{normal}
\eea
where $\overline{\Omega(0, \mu_0, \xi)} = \Omega_{cl}(\xi)$ plus divergent vacuum terms. These are reminiscent of the usual 
normalization conditions employed in the effective potential approach of quantum field theories.

In both expressions for $E_0$ and $\Pi_v$ we have terms proportional to $\ln\left(1+4\lambda\xi^2/m^2\right)$ which could render 
the renormalization procedure somewhat cumbersome. Since $\lambda \xi \ll 1$ by assumption, for simplicity we may Taylor expand 
this logarithmic function and keep terms up to $\lambda^2\xi^4$. Using this technique for $E_0$ and $\Pi_v$ and adding the resulting 
divergent term $E_0 + m^4\sigma^2\Pi_v^2/2$ to the classical energy density~(\ref{cla-therm}) results in the following vacuum energy density:
\begin{widetext}
\bea
&&\overline{\Omega(0, \mu_0, \xi)} =\frac{1}{1024\pi^4}\Biggl\{ 64 \pi^2 m^2 \Lambda_c^2 - 16\pi^2 m^4 
- 32 \pi^2 m^4 \ln\left(\frac{\Lambda_c^2}{m^2}\right)
\nn\\
&&+\, m^4 \left[\ln\left(\frac{\Lambda_c^2}{m^2}\right)\right]^2 +\frac{m^4\sigma^2 \Lambda_c^2}{2}\left[I(m,\Lambda_c)  -
 m^2 \ln \left(\frac{\Lambda_c^2}{m^2}\right)\right] \Biggr\}
\nn\\
&&+\, \xi ^2 \left\{ m^2 - \mu_0^2 + \delta m^2 +\frac{\lambda}{4 \pi ^2}\,I(m,\Lambda_c) 
- \frac{\lambda m^2}{128 \pi ^4}\ln\left(\frac{\Lambda_c^2}{m^2}\right)  + \frac{m^4\sigma^2}{2}\left[g(\lambda,m,\Lambda_c) 
- \frac{\lambda\Lambda_c^2}{128 \pi ^4}\ln\left(\frac{\Lambda_c^2}{m^2}\right)\right] \right\}
\nn\\
&&+\,\xi ^4 \left\{ \lambda +\delta \lambda + f_{+}(\lambda) + \ln\left(\frac{\Lambda_c^2}{m^2}\right)f_{-}(\lambda)
+ m^4\sigma^2\left[\frac{\lambda}{m^2}\,g(\lambda,m,\Lambda_c) - \frac{\lambda^2}{32 \pi^4}\ln \left(\frac{\Lambda_c^2}{m^2}\right)\right] \right\},
\label{vac-div}
\eea
where $\delta m^2 \xi^2$ and $\delta\lambda\xi^4$ are the above mentioned counterterms and
\bea
&&f_{\pm}(\lambda) = \frac{\lambda ^2}{2\pi ^2}\left(\frac{1}{32 \pi ^2} \pm 1\right),
\nn\\
&& g(\lambda,m,\Lambda_c) = \frac{\lambda\Lambda_c^2}{128 \pi ^4}+ \frac{\lambda m^2}{128\pi ^4}\,\left[\ln\left(\frac{\Lambda_c^2}{m^2}\right)\right]^2,
\nn\\
&& I(m,\Lambda_c) = \Lambda_c^2 - m^2\,\ln\left(\frac{\Lambda_c^2}{m^2}\right).
\eea
In Eq.~(\ref{vac-div}) we again have retained terms up to $\lambda^2\xi^4$. Employing the normalization conditions~(\ref{normal}) we find that
\beq
\delta m^2 =  \frac{\lambda m^2}{128 \pi ^4}\,\ln\left(\frac{\Lambda_c^2}{m^2}\right) - \frac{\lambda}{4 \pi ^2}\,I(m,\Lambda_c)  
+ \frac{m^4\sigma^2}{2}\left[\frac{\lambda \Lambda_c^2}{128 \pi^4}\,\ln\left(\frac{\Lambda_c^2}{m^2}\right) - g(\lambda,m,\Lambda_c)\right],
\eeq
and
\beq
\delta \lambda = - f_{+}(\lambda) - \ln\left(\frac{\Lambda_c^2}{m^2}\right)f_{-}(\lambda) -  m^4\sigma^2\left[\frac{\lambda}{m^2}\,g(\lambda,m,\Lambda_c) 
- \frac{\lambda^2}{32 \pi^4}\ln \left(\frac{\Lambda_c^2}{m^2}\right)\right].
\eeq
In this way, after a straightforward calculation one gets
\beq
\overline{\Omega(0, \mu_0, \xi)} = \Omega_{cl}(\xi) + K(\lambda,m,\Lambda_c),
\label{vac-div2}
\eeq
where
\beq
K(\lambda,m,\Lambda_c) = -\frac{m^4}{64 \pi ^2}+\frac{m^2}{32 \pi ^2}\left\{\Lambda_c^2 + I(m,\Lambda_c)
+ \frac{m^2}{32 \pi ^2}\,\left[\ln\left(\frac{\Lambda_c^2}{m^2}\right)\right]^2  \right\}
+\frac{m^4\sigma^2\Lambda_c^2}{2048\pi^4}\left[I(m,\Lambda_c) - m^2\,\ln\left(\frac{\Lambda_c^2}{m^2}\right)\right].
\eeq
The (infinite) constant term $K(\lambda,m,\Lambda_c)$ can be set to zero by shifting the vacuum energy density by a constant amount. 
This can always be done since in non-gravitational physics only energy differences are measurable. In this way, we finally get that 
the renormalized vacuum energy is just the classical energy density, $\Omega_{cl}(\xi) =\overline{\Omega(0, \mu_0, \xi)}$.
\end{widetext}

\section{Ultrarelativistic limit of $\Pi_m(\beta,\mu, \xi)$}
\label{app5}

Employing spherical coordinates, one can express $\Pi_m(\beta, \mu, \xi)$ as
\bea
\Pi_m(\beta, \mu, \xi) &=& \frac{1}{4\pi^2}\int_{0}^{\infty}\,dp\,\frac{p^2}{\omega}\biggl[\frac{1}{e^{\beta(\omega-\mu)}-1}
\nn\\
&+&\frac{1}{e^{\beta(\omega+\mu)}-1}\biggr],
\eea
where a partial integration was made and $\omega = \omega(\p)$. Let us define
\bea
g_l(y,r) &=& \frac{1}{\Gamma(l)}\,\int_{0}^{\infty}dx
\nn\\
&\times&\,\frac{x^{l-1}}{\exp{[(x^2+y^2)^{1/2} - r y]}-1},
\eea
\bea
h_l(y,r) &=& \frac{1}{\Gamma(l)}\,\int_{0}^{\infty}\frac{dx}{(x^2+y^2)^{1/2}}
\nn\\
&\times&\,\frac{x^{l-1}}{\exp{[(x^2+y^2)^{1/2} - r y]}-1}.
\eea
The functions of interest here are
\beq
G_l(y,r) = g_l(y,r) - g_l(y,- r),
\eeq
\beq
H_l(y,r) = h_l(y,r) + h_l(y,- r).
\eeq
Therefore, with $y = \beta M$, $r = \mu/M$ and after a simple change of variables we get
\beq
\Pi_m(y, r) = \frac{1}{2\pi^2\beta^2}H_3(y,r).
\label{pim-h3}
\eeq
The calculation of the functions $G_l$ and $H_l$ is discussed at length in Ref.~\cite{haber}. Here we simply 
quote the quantities which are relevant for our computations. The following recursion relations ought to be 
employed:
\beq
\frac{d G_{l+1}}{dy} = lrH_{l+1} - \frac{y}{l}\,G_{l-1} + \frac{y^2 r}{l}\,H_{l-1},
\label{haber1}
\eeq
\beq
\frac{d H_{l+1}}{dy} = \frac{r}{l}\,G_{l-1} - \frac{y}{l}\,H_{l-1},
\label{haber2}
\eeq
with the initial conditions $G_l(0,0) = 0$, $l > 0$, and $H_l(0,0) = 2\zeta(l-1)/(l-1)$, $l > 2$, $\zeta(s)$ being the usual 
Riemann zeta function. Consequently, knowledge of $G_1$ and $H_1$ will yield $G_l$ and $H_l$ for all positive odd $l$.

The small $y$ expansions of the functions $G_1$ and $H_1$ are given by, respectively:
\bea
G_1(y,r) &=& \frac{\pi r}{(1-r^2)^{1/2}} - r y + 2\pi r
\nn\\
&\times&\,\sum_{k=1}^{\infty}\,(-1)^{k+1}\,a_k\,\zeta(2k+1)\left(\frac{y}{2\pi}\right)^{2k+1},
\label{haber3}
\eea
and
\bea
H_1(y,r) &=& \frac{\pi}{y(1-r^2)^{1/2}} + \ln\left(\frac{y}{4\pi}\right) + \gamma
\nn\\
&+&\,\sum_{k=1}^{\infty}\,(-1)^{k}\,b_k\,\zeta(2k+1)\left(\frac{y}{2\pi}\right)^{2k},
\label{haber4}
\eea
with $\gamma = 0.5772...$ being the Euler's constant. The quantities $a_k$ and $b_k$ are simple polynomials in $r$. For $k=1$ one 
has $a_1 = 1$ and $b_1 = r^2 + 1/2$. We refer the reader to~\cite{haber} for all important details concerning the derivations of 
the above relations.

The $y \ll 1$ limit allows retain just the first term of the summations in $G_1$ and $H_1$. 
Employing Eqs.~(\ref{haber1}),~(\ref{haber2}),~(\ref{haber3}) and~(\ref{haber4}) with the aforementioned initial conditions 
one obtains, after a straightforward calculation:
\bea
H_3(y,r) &=& \frac{\,\pi^2}{6} - \frac{y}{2}\,\pi(1 - r^2)^{1/2}
\nn\\
&+& \frac{y^2}{8}\left[\ln\left(\frac{16\pi^2}{y^2}\right) -2\gamma + 1 - 2r^2\right]
\nn\\
&+& \frac{y^4}{64\pi^2}\,(1+ 4 r^2)\,\zeta(3),
\label{h3}
\eea
where we used the fact that $\zeta(2) = \pi^2/6$. Hence inserting the above expressions in Eq.~(\ref{pim-h3}) we get
\beq
\Pi_m(\beta, \mu, \xi) \approx \frac{1}{12 \beta ^2}-\frac{\mu^2}{8 \pi ^2}+\frac{M^2}{16 \pi^2}B_1(\beta),
\label{pim58}
\eeq
where
\beq
B_1(\beta) = \ln \left(\frac{16\pi^2}{\beta^2  M^2}e^{- 2\gamma +1}\right).
\label{B}
\eeq
Terms proportional to $\beta^2$ or higher powers of $\beta$ were dropped.

\end{document}